\documentstyle[twoside,fleqn,espcrc2]{article}

\newcommand{\AmS}{{\protect\the\textfont2
  A\kern-.1667em\lower.5ex\hbox{M}\kern-.125emS}}

\title{Progress and status of APEmille}

\def\r{$^{\rm a}$}
\def\p{$^{\rm c}$}
\def\z{$^{\rm d}$}
\author{A.~Bartoloni\address{INFN, Sezione di Roma I, I-00100 Roma (Italy)}, 
        S.~Cabasino\r, 
        N.~Cabibbo\address{ENEA, I-00100 Roma (Italy),
                           and INFN, Sezione di Roma I, I-00100 Roma (Italy)}, 
        M.~Cosimi\r, 
        P.~De Riso\r,
        W.~Errico\address{INFN, Sezione di Pisa, 
                               I-56010 San Piero a Grado (Italy)},
        S.~Giovannetti\p,
        F.~Laico\p, 
        H.~Leich\address{DESY, Platanenallee 6, 15636 Zeuthen (Germany)}, 
        A.~Lonardo\r, 
        G.~Magazz\'u\p, 
        A.~Michelotti\r, 
        E.~Panizzi\r, 
        P.S.~Paolucci\r,
        D.~Rossetti\r, 
        U.~Schwendicke\z,
        H.~Simma\z\thanks{speaker at the conference (e-mail: simma@ifh.de)},
        K.H.~Sulanke\z, 
        M.~Torelli\r, 
        R.~Tripiccione\p,
        P.~Vicini\r
}

\begin{document}

\begin{abstract}
We report on the progress and status of the APEmille project: a SIMD 
parallel computer with a peak performance in the TeraFlops range 
which is now in an advanced development phase.
We discuss the hardware and software architecture,
and present some  performance estimates for Lattice Gauge Theory (LGT) 
applications. 
\end{abstract}

\maketitle

\section{Introduction}
The APE project is one of a small number of physicist-driven efforts
in Europe, Japan \cite{uka}  and the US \cite{arsenico},
aiming at the development of number crunchers
optimised for LGT applications.
APEmille, the latest generation of the APE family, 
is an evolution of the APE100 architecture \cite{a100} and has 
a roughly one order of magnitude higher peak performance.

APEmille is a versatile massive-parallel processor optimised for 
heavy duty numerical applications and scalable from 4 GigaFlops 
to 1 TeraFlops.
The main architectural enhancements over the previous generation
are the availability
of local addressing on each node, the support of floating-point (FP)
arithmetics in double precision and for complex data types, and 
the realisation of a multiple SIMD architecture. 
It allows to divide the machine into independent partitions running
different programs and one program can also activate
multiple SIMD threads. These enhancements make the machine 
potentially useful in a broader class of applications that require 
massive numerical computations and enjoy some degree of locality, 
e.g. complex systems and fluid dynamics.

In this paper, we describe some details of the APEmille architecture
and the status of its hardware implementation. We also discuss main 
software elements and performance aspects.

\section{Architecture and Hardware}
APEmille is organised as a three-dimensional array of
nodes, each consisting of an arithmetic processor 
and its local data memory. 
Each group of eight nodes is controlled by 
a control processor and operates in SIMD mode
with local addressing: the nodes synchronously execute the same 
instructions and may access their memories using different locally
calculated addresses.
The smallest machine partition that can run a program independently 
has eight nodes. Larger SIMD partitions are obtained by running an
identical instruction stream on the corresponding control processors.
The arithmetic processors can directly access the data memories of 
remote nodes by means of a flexible communication network.

The hardware building-block of APEmille is a processing board (PB). 
It houses eight nodes (logically arranged at the corners of a cube),
one control processor, and a communication controller. 
In addition, each PB has its own PCI-based host-interface.
A large APEmille machine is assembled by replicating 
these PB's (connected by communication-lines and global control-signals).
The replication of the control processor and host interface 
on each board not only allows simple software configuration of 
the machine into multiple SIMD partitions, but also provides
a highly simplified and modular system-level hardware with 
improved reliability of the global system.

The control processor (called ``Tarzan'') has its own memory for
data (512 kByte SRAM) and program (512 k instructions DRAM). 
Tarzan manages the program flow (branches, loops, 
subroutine and function calls etc.) and generates the global memory 
address for the arithmetic processors.
Tarzan contains two separate units with short pipelines for integer 
arithmetics and address generation. They use a large register file 
(256 words \`a 32 bit) and provide special instructions e.g. for efficient 
loop control (increment, comparison and conditioned branch) within a 
single clock cycle.

The arithmetic processor (called ``Jane'') supports several data 
types: Floating-point {\em normal}
operations ($a\times b + c$) are implemented
for 32- and 64-bit IEEE format, as well as for single precision complex 
and vector (pairs of 32-bit) operands.
Arithmetic and bit-wise operations are available for integer data types.
Operands can be converted between the various formats.
The combined double pipeline of Jane allows to start one arithmetic 
operation in every clock cycle (with pipeline length of 9, as in APE100, 
or less). This corresponds to two, four, or eight Flop per cycle
for normal operations with simple (SNORM or DNORM), 
vector (VNORM) or complex (CNORM) operands, respectively.
The key element for an efficient filling of the arithmetic pipeline
is the large register file with 512 words \`a 32 bits, which avoids
the need for an intermediate cache between memory and registers.

Jane manages its own local data memory, a synchronous DRAM with 4 Mwords
\`a 64 bit. 
The local address generation unit, which is independent of the arithmetic 
operations of Jane, combines the global address received from Tarzan with 
a local address offset on the node. 
The memory is organised in two interleaved banks to allow page-fault free 
access for data bursts up to the size of the entire register file. 
The bandwidth for the transfer between memory and registers is one (two) 
operands \`a 64 (32) bits per clock cycle and start-up latencies between 
subsequent bursts can  be reduced to a few cycles by partial overlapping.

Memory access to remote nodes is automatically handled by the communication 
network. It provides homogeneous communications over arbitrary distances,
i.e. all nodes can simultaneously access data from a corresponding remote node 
with a given relative distance. 
The communication network automatically exploits multiple data paths 
for communications along more than one coordinate axes and it allows broadcast 
over the full machine, as well as along lines and planes of nodes. 
``Soft'' or inhomogeneous communications, where the relative distance to 
the remote node may be different on each node, are forseen as a future 
extension.

Tarzan, Jane, and the communication controller are built into
three custom designed ASIC circuits with very low power 
consumption. The APEmille chips will probably be built with two 
successive technologies, increasing the clock frequency. The first
APEmille systems
will operate at 66 MHz providing 500 MFlops per processor. 
Upgraded to 100 MHz, the full machine with 2048 nodes would then deliver 
a peak performance of 1.6 TeraFlops.

APEmille is hosted by a cluster of networked PCI-based PC's or workstations. 
Each host controls a group of 32 memory-mapped nodes.
The hosts are connected by a dedicated high-speed low-latency network 
based on serial Motorola ``AutoBahn'' links \cite{desy} or by any standard 
network interface.
The close integration of APEmille into a host network allows a high I/O
bandwidth and flexible re-configuration of the machine and the peripherals.

\section{Software and Simulation Environment}

The software developed for APEmille has three main parts, the
operating system (OS), the compilation chain (TAOmille), and development 
and simulation tools.

The OS is distributed on the hosts running a freely available Linux Kernel.
The OS handles the program loading, local and global data I/O, system services,
and the partitioning of the machine.
The OS has a layered and object-oriented structure written 
in C++, and uses an innovative client server approach which supports the 
creation and management of distributed objects.

The dynamic programming language TAOmille supports the new features of 
the architecture (new data types, local addressing, extended communications)
but maintains full compatibility with the language of APE100.
An important new element of the compiler is a high-level optimisation stage
before the assembler code generation. It eliminates e.g. common 
subexpressions and performs optimisations for efficient address 
calculations and memory access. 
A modular structure of the compilation chain allows future extensions
to other programming languages by changing only the uppermost level of 
the compiler.

A freely configurable simulator written in C++ is available for 
detailed behavioural simulations of all APEmille components.
It is used for automatic generation of test stimuli and reference 
outputs for the electrical VHDL simulation, for tests of the
compilation chain and OS, and for performance estimates
of realistic application codes.

\section{Performance Considerations}

Tarzan as well as Jane use a very long instruction word, in which the 
operations of all independent devices are specified at each clock cycle. 
This allows a highly efficient scheduling and optimisation of the 
arithmetic and memory-access pipelines by the low-level optimiser in 
the last step of the compilation chain.

Various new controller instructions and their improved temporisation
allow in general shorter latencies for address calculations and
should improve related performance bottle-necks in APE100. The local 
addressing and the more flexible communications will be helpful
for various algorithmic problems, like preconditioning or FFT.

Maximal FP performance can be obtained in APEmille by the use of
complex normal operations (8 Flop/cycle) for which the ratio between
Flop rate and memory bandwidth is as well-balanced as in APE100. 
While for other operand types the maximal Flop rate 
of Jane is smaller (1/2 for VNORM, 1/4 for DNORM and SNORM), the 
correspondingly increased bandwidth/Flop rate can be helpful for 
applications that are less balanced than LGT computations.

Rough performance estimates with simple and un-tuned assembler code
for the Wilson-Dirac operator yield 88\% pipeline filling (760 clock 
cycles) when using CNORM instructions, corresponding to 60\% peak 
performance (subtracting useless Flops from e.g. real$\times$complex 
factors). This shows that no severe limitations from the memory 
bandwidth are encountered. For double precision a pipeline filling
of 82\% (1150 clock cycles) has been found, corresponding to 21\%
of single-precision peak performance. This indicates that less than
the naive factor of 4 is lost with double precision arithmetics.

\section{Status Summary}

The main elements of the software are available and tested on 
the system simulator: A prototype of the distributed OS is 
running, and the low level of the new compilation chain 
is used extensively for the compilation of test programs.

The detailed electronic design of APEmille is almost completed 
and some VLSI components are ready to be produced.
Extensive tests of the overall design are presently performed on
the simulator. A register-file test chip is the first VLSI 
component available so far, and further hardware components of the 
system will be tested during the next months. 
We expect to have first prototype systems with several APEmille 
boards early in spring next year and a medium size machine 
for the next summer.

\end{document}